\documentclass[reprint,
superscriptaddress,amsmath,amssymb,aps, longbibliography,nofootinbib
]{revtex4-2}

\usepackage{graphicx}
\usepackage{dcolumn}
\usepackage{bm}
\usepackage{siunitx}
\usepackage[noend]{algorithmic}
\usepackage{algorithm}
\usepackage[skins]{tcolorbox}
\usepackage{setspace}
\usepackage{blkarray}

\usepackage{xcolor}
\usepackage{soul}
\usepackage{systeme}
\usepackage{footmisc}
\usepackage{mhchem}

\usepackage{scalerel} 

\makeatletter
\def\maketag@@@#1{\hbox{\m@th\normalfont\normalsize#1}}
\makeatother

\usepackage{booktabs}
\usepackage{hyperref}
\usepackage[capitalize]{cleveref}
\usepackage[utf8]{inputenc}

\usepackage[lining,semibold]{libertine} 
\usepackage{amsthm}
\usepackage[libertine, cmintegrals, bigdelims, vvarbb]{newtxmath}

\usepackage{amsmath}
\usepackage{amsfonts}
\usepackage{mathrsfs}
\usepackage{gensymb}
\usepackage{bbm}
\usepackage{dsfont}

\usepackage{pgfplots}
\usepgfplotslibrary{colormaps}

\usepackage{kbordermatrix}
\renewcommand{\kbldelim}{(}
\renewcommand{\kbrdelim}{)}

\usepackage{chemformula}
\usepackage[caption=false]{subfig}

\usepackage{soul}
\usepackage{xcolor}

\theoremstyle{definition}

\usepackage{scalerel} 

\definecolor{webgreen}{rgb}{0,.5,0}
\definecolor{webbrown}{rgb}{.6,0,0}
\definecolor{grigio}{rgb}{.85,.85,.85} 
\definecolor{RoyalBlue}{rgb}{0.0, 0.14, 0.4}
\definecolor{skyblue1}{rgb}{0.45,0.62,0.81}
\definecolor{skyblue2}{rgb}{0.2,0.39,0.64}
\definecolor{skyblue3}{rgb}{0.13,0.29,0.53}
\definecolor{scarlet1}{rgb}{0.93,0.16,0.16}
\definecolor{scarlet2}{rgb}{0.8,0,0}
\definecolor{scarlet3}{rgb}{0.64,0,0}

\definecolor{g}{gray}{0.50}

\hypersetup{%
    colorlinks=true, linktocpage=true, pdfstartpage=1, pdfstartview=FitV,%
    breaklinks=true, pdfpagemode=UseNone, pageanchor=true, pdfpagemode=UseOutlines,%
    plainpages=false, bookmarksnumbered, bookmarksopen=true, bookmarksopenlevel=1,%
    hypertexnames=true, pdfhighlight=/O,
    urlcolor=webbrown, linkcolor=RoyalBlue, citecolor=webgreen, 
    pdftitle={},%
    pdfauthor={Timur Aslyamov},%
    pdfsubject={},%
    pdfkeywords={},%
    pdfcreator={pdfLaTeX},%
    pdfproducer={LaTeX REVTeX}%
}

\DeclareFontFamily{U}{BOONDOX-calo}{\skewchar\font=45 }
\DeclareFontShape{U}{BOONDOX-calo}{m}{n}{
  <-> s*[1.05] BOONDOX-r-calo}{}
\DeclareFontShape{U}{BOONDOX-calo}{b}{n}{
  <-> s*[1.05] BOONDOX-b-calo}{}
\DeclareMathAlphabet{\mathcalboondox}{U}{BOONDOX-calo}{m}{n}
\SetMathAlphabet{\mathcalboondox}{bold}{U}{BOONDOX-calo}{b}{n}
\DeclareMathAlphabet{\mathbcalboondox}{U}{BOONDOX-calo}{b}{n}

\begin{document}
\title{Nonequilibrium fluctuation-response relations for state-current correlations}

\author{Krzysztof Ptaszy\'{n}ski}
\email{krzysztof.ptaszynski@ifmpan.poznan.pl}
\affiliation{Institute of Molecular Physics, Polish Academy of Sciences, Mariana Smoluchowskiego 17, 60-179 Pozna\'{n}, Poland}

\author{Timur Aslyamov}
\email{timur.aslyamov@uni.lu}
\affiliation{Complex Systems and Statistical Mechanics, Department of Physics and Materials Science, University of Luxembourg, 30 Avenue des Hauts-Fourneaux, L-4362 Esch-sur-Alzette, Luxembourg}

\author{Massimiliano Esposito}
\email{massimiliano.esposito@uni.lu}
\affiliation{Complex Systems and Statistical Mechanics, Department of Physics and Materials Science, University of Luxembourg, 30 Avenue des Hauts-Fourneaux, L-4362 Esch-sur-Alzette, Luxembourg}

\date{\today}

\begin{abstract}
Recently, novel exact identities known as Fluctuation-Response Relations (FRRs) have been derived for nonequilibrium steady states of Markov jump processes. These identities link the fluctuations of state or current observables to a combination of responses of these observables to perturbations of transition rates. Here, we complement these results by deriving analogous FRRs applicable to mixed covariances of one state and one current observable. We further derive novel Inverse FRRs expressing individual state or current response in terms of a combination of covariances rather than vice versa. Using these relations, we demonstrate that the breaking of the Onsager symmetry requires the presence of state-current correlations. On the practical side, we demonstrate the applicability of FRRs for explaining the behavior of fluctuations in quantum dot devices or enzymatic reaction networks.
\end{abstract}

\maketitle

\section{Introduction}
The dynamical behavior of small systems, relevant in fields as diverse as biochemistry or nanoelectronics, is intrinsically stochastic, that is, characterized by large fluctuations around the average behavior. The statistical description of this stochastic behavior can be provided by analyzing the ensemble of stochastic trajectories of the system. Within the field of nonequilibrium statistical physics, one often considers two classes of trajectory-based observables. The first are time-integrated state observables corresponding, e.g., to a fraction of time spent by the system in a given state or a pool of states. They are the focus of a companion Letter~\cite{ptaszynski2024nonequilibrium}. The second ones are time-integrated currents that are expressed in terms of the number of transitions between the system states~\cite{aslyamov2024frr}. Physically, such currents may correspond to the exchange of some quantity (e.g., electric charge or heat) with the reservoir, number of steps of the molecular motor, etc. Although the state and current observables are often considered separately, the properties of their joint distribution attracted attention in the context of electronic transport~\cite{utsumi2007full,utsumi2007fullb,utsumi2008full} and diffusion~\cite{dieball2022correlations}. In particular, it has been shown that in the long-time limit of Markovian dynamics, the covariances of state and current observables vanish at equilibrium due to the time-reversal symmetry~\cite{dieball2022correlations}. Therefore, their presence indicates the nonequilibrium nature of the system.

At equilibrium, fluctuations of observables are related to their responses to external perturbations by a seminal fluctuation-dissipation theorem~\cite{kubo1966fluctuation,kubo2012statistical,stratonovich2012nonlinear,marconi2008fluctuation,forastiere2022linear}. Away from equilibrium, this theorem does not hold, although certain generalizations to the nonequilibrium regime have been proposed~\cite{agarwal1972fluctuation,baiesi2009fluctuations,seifert2010fluctuation,prost2009generalized,altaner2016fluctuation,maes2020response,chun2023trade,shiraishi2023introduction,gao2024thermodynamic,tesser2024out}. Nevertheless, research in recent decades has produced a wealth of universal laws describing the properties of fluctuations (e.g., fluctuation theorems~\cite{jarzynski1997nonequilibrium,crooks1999entropy,esposito2009nonequilibrium,seifert2012stochastic} and thermodynamic or kinetic uncertainty relations~\cite{barato2015thermodynamic,gingrich2016dissipation,pietzonka2016universal, pietzonka2017finite,horowitz2017proof,falasco2020unifying,horowitz2020thermodynamic,vu2020entropy,van2023thermodynamic,di2018kinetic,shiraishi2021optimal}) and responses~\cite{lucarini2016response,santos2020response,falasco2019negative, mallory2020kinetic,owen2020universal,owen2023size,gabriela2023topologically,aslyamov2024nonequilibrium,aslyamov2024general,harunari2024mutual,cengio2025mutual,khodabandehlou2024affine,floyd2024learning,frezzato2024steady,floyd2024limits,gao2022thermodynamic,bao2024nonequilibrium,zheng2025spatial} in Markov processes or chemical reaction networks. Significant developments have also been made in connecting responses and fluctuations, in the spirit of the original fluctuation-dissipation theorem, in systems arbitrarily far from equilibrium~\cite{dechant2019arxiv,dechant2020fluctuation,falasco2022beyond,monnai2024kinetic}. In particular, Refs.~\cite{zheng2024information,ptaszynski2024dissipation,aslyamov2024frr,liu2024dynamical,kwon2024fluctuation,van2024fundamental,liu2025response} derived novel inequalities bounding the precision of response (i.e., their ratio to fluctuations) of trajectory-based observables in terms of entropy production rate or traffic (activity), a quantity that measures the total number of transitions per unit time in the system.

Going beyond inequalities, Refs.~\cite{aslyamov2024frr,ptaszynski2024nonequilibrium} derived exact identities, called Fluctuation-Response Relations (FRRs), relating fluctuations and responses of current or state observables in a Nonequilibrium Steady State (NESS) of Markov jump processes. More precisely, these identities express the covariance of two current or state observables in terms of combination of the responses of these observables to perturbations of the transition rates. Here, we complement this result by deriving analogous FRRs applicable to mixed covariances of state and current observables. These relations are further generalized to correlations between state and generic flux observables, which recently attracted some attention~\cite{bakewell2023general,pietzonka2024thermodynamic}. We also derive inverse relations, called \textit{Inverse FRRs}, expressing a state or current response to a single perturbation in terms of the covariances of state and current observables. One of the consequences of these relations is that the breaking of the Onsager symmetry requires the presence of the state-current covariances. On the practical side, we show how our result can be used to gain physical insight into the behavior of stochastic systems relevant for electronic transport (quantum dots) or biochemistry (enzymatic reactions).

The paper is organized as follows. In Sec.~\ref{sec:setup} we describe our theoretical framework. In Sec.~\ref{sec:observables} we define the state and current observables. In Sec.~\ref{sec:frrs} we present the obtained FRRs and Inverse FRRs. In Secs.~\ref{sec:calc-fluct}--\ref{sec:examples} we discuss how FRRs can be used to calculate fluctuations and gain physical insight into the behavior of fluctuations in electronic transport or enzymatic reaction networks. Finally, in Sec.~\ref{sec:concl} we draw the conclusions. In the Appendices we present the derivation of the formula for the state-current covariance, the proof of our main result, and some additional derivations and expressions.

\section{Framework} \label{sec:setup}
We consider a continuous-time Markov jump process among $N$ discrete states. It is described by the graph whose nodes correspond to the system states and the undirected edges $e$ to the transitions between states. We further make the graph directed by assigning each edge $e$ a forward ($+e$) and reverse ($-e$) direction, so that the source of the directed edge $\pm e$, labeled $s(\pm e)$, is a target of the directed edge $\mp e$, labeled  $t(\mp e)$. The transition rate associated with the directed edge $\pm e$ is denoted as $W_{\pm e}$. The NESS of the system is defined by
\begin{align}
\label{eq:NESS}
    d_t\boldsymbol{\pi} = \mathbb{W}\cdot\boldsymbol{\pi} = 0 \,,
\end{align}
where $\boldsymbol{\pi}=(\dots,\pi_n,\dots)^\intercal$ is the vector of state probabilities $\pi_n$ with $\sum_n\pi_n=1$. The matrix $\mathbb{W}$ is the rate matrix with off-diagonal elements $W_{nm}=\sum_{e}[ W_{+e}\delta_{s(+e)m}\delta_{t(+e)n} + W_{-e}\delta_{s(-e)m}\delta_{t(-e)n}]$, where $\sum_e$ denotes the summation over undirected edges $e$, and the diagonal elements $W_{nn}=-\sum_{m \neq n}W_{mn}$. 

We further employ a generic parameterization of the transition rates~\cite{owen2020universal,aslyamov2024nonequilibrium}
\begin{align}
\label{eq:rates-model}
    W_{\pm e}=\exp(B_e \pm S_e/2) \,,
\end{align}
where $B_e$ and $S_e$ parametrize the symmetric and antisymmetric parts of the transition rate, respectively. 
For physical systems in contact with thermal reservoirs (rates that satisfy the local detailed balance), the term $B_e$ characterizes the kinetic barrier between the system states. Physically, it can be controlled by varying catalyst (e.g., enzyme) concentrations~\cite{Wachtel_2018}, applying magnetic fields (e.g., via the radical pair mechanism in magnetoreception)~\cite{ritz2000model, hore2016radical, wiltschko2019magnetoreception, zadeh2022magnetic}, or adjusting tunnel barriers~\cite{gustavsson2006counting,gustavsson2009electron,sanchez2019Autonomous} or potential barriers~\cite{freitas2021stochastic, gopal2022Large} by gate voltages in nanoelectronics. The term $\pm S_e$, instead, is the change in entropy in the reservoir due to a transition along the edge $\pm e$ that includes changes in thermodynamic forces and the energy landscape \cite{rao2018conservation, falasco2023macroscopic, owen2020universal}.

\section{State and current observables} \label{sec:observables}
Our object of interest are two kinds of random variables, time-integrated state and current observables. The state observables are defined as
\begin{align} \label{eq:stateobsdef}
\hat{o}(t) \equiv \frac{1}{t} \sum_{n} o_n \int_{0}^t \phi_n(t') dt' \,,
\end{align}
where the integral is performed over a stochastic trajectory of the system. Here, $\boldsymbol{o} \equiv (\dots, o_n, \dots)^\intercal$ is the vector that defines the observable and $\phi_n(t)$ is the random variable taking the value $1$ when the state $n$ is occupied and $0$ otherwise. Analogously, time-integrated current observables are defined as
\begin{align} \label{eq:curobsdef}
\hat{J}(t) \equiv \frac{1}{t} \sum_e x_e  [k_{+e}(t)-k_{-e}(t) ]\,,
\end{align}
where $\boldsymbol{x} \equiv (\ldots,x_e,\ldots)$ and $k_{\pm e}(t)$ is the number of jumps along  directed edge $\pm e$ during the time interval $[0,t]$. We note that current observables are time-antisymmetric because jumps along forward and backward edges $+e$ and $-e$ bring contributions with the same magnitude $|x_e|$ but the opposite sign. In Sec.~\ref{subsec:fluxobs} we will consider a generalization of our framework to generic flux observables without that property. 

The average values of these observables are defined as
\begin{align} \label{eq:stateobsmean}
\mathcal{O} &\equiv \lim_{t \rightarrow \infty} \langle \hat{o}(t) \rangle =\sum_n o_n \pi_n \,, \\
\mathcal{J} &\equiv \lim_{t \rightarrow \infty} \langle \hat{J}(t) \rangle =\sum_e x_e j_e \,,
\end{align}
where $\langle \cdot \rangle$ denotes the average over the ensemble of stochastic trajectories and $j_e \equiv W_{+e} \pi_{s(+e)}-W_{-e} \pi_{t(+e)}$ is the directed current along the edge $e$. The covariance of state and current observables is defined as
\begin{align}
\langle \langle \mathcal{O},\mathcal{J} \rangle \rangle \equiv \lim_{t \rightarrow \infty} t \langle \Delta \hat{o}(t) \Delta \hat{J}(t) \rangle \,,
\end{align}
where $\Delta \hat{o}(t) \equiv \hat{o}(t)-\langle \hat{o}(t) \rangle$ and $\Delta \hat{J}(t) \equiv \hat{J}(t)-\langle \hat{J}(t) \rangle$. It is given by the algebraic expression
\begin{align}
\langle \langle \mathcal{O},\mathcal{J} \rangle \rangle=\boldsymbol{o}^\intercal \mathbb{C}^\mathfrak{m} \boldsymbol{x} \,,
\end{align}
where $\mathbb{C}^\mathfrak{m}=[C^\mathfrak{m}_{ne}]$ is the covariance matrix with elements defined as
\begin{align} \label{eq:covmat-def}
C^\mathfrak{m}_{ne} \equiv \lim_{t \rightarrow \infty} t^{-1} \big \langle \theta_n(t) [\Delta k_{+e}(t)-\Delta k_{-e}(t)] \big \rangle \,,
\end{align}
where $\theta_n(t) \equiv \int_0^t [\phi_n(t')-\pi_n] dt'$ and $\Delta k_{\pm e}(t) \equiv k_{\pm e}(t)-\langle k_{\pm e}(t) \rangle$. Here, with the superscript $\mathfrak{m}$ we denote the ``mixed'' state-current covariances. The elements of the covariance matrix can be calculated using a tilted rate matrix~\cite{utsumi2007full} $\mathbb{W}^\phi(\boldsymbol{q},\boldsymbol{\zeta})$ with off-diagonal elements 
$W_{nm}=\sum_{e}[ W_{+e}\delta_{s(+e)m}\delta_{t(+e)n} e^{q_e} + W_{-e}\delta_{s(-e)m}\delta_{t(-e)n} e^{-q_e}]$
and diagonal elements $W_{nn}^\phi(\boldsymbol{q},\boldsymbol{\zeta}) =W_{nn}+\zeta_n$. Explicitly, they are given by the expression (see Appendix~\ref{app:der-covmat-expression})
\begin{align} \label{eq:covmat-expression}
&C^\mathfrak{m}_{ne}=-\boldsymbol{1}^\intercal \mathbb{J}_e \mathbb{W}^D \mathds{1}^{(n)} \boldsymbol{\pi}-\boldsymbol{1}^\intercal \mathds{1}^{(n)} \mathbb{W}^D \mathbb{J}_e \boldsymbol{\pi} \,,
\end{align}
where $\boldsymbol{1}=(\ldots,1,\ldots)^\intercal$ is the vector of ones and $\mathbb{W}^D$ is the Drazin inverse of the rate matrix, which has been generically defined in Ref.~\cite{drazin1958pseudo}. In our context, it corresponds to a unique solution of the equation $\mathbb{W} \mathbb{W}^D=\mathds{1}-\boldsymbol{\pi} \boldsymbol{1}^\intercal$. See Refs.~\cite{crook2018drazin,landi2023current,bao2024nonequilibrium,harvey2023universal} for more of its properties and applications characterizing fluctuations and responses. The matrices
\begin{subequations} \label{eq:curprojop}
    \begin{align}
    \mathbb{J}_e &\equiv \frac{\partial}{\partial q_{e}} \mathbb{W}^\phi(\boldsymbol{q},\boldsymbol{\zeta})\Big|_{\boldsymbol{q},\boldsymbol{\zeta}=\boldsymbol{0}}\,, \\
     \mathds{1}^{(n)} &\equiv \frac{\partial}{\partial \zeta_{n}} \mathbb{W}^\phi(\boldsymbol{q},\boldsymbol{\zeta})\Big|_{\boldsymbol{q},\boldsymbol{\zeta}=\boldsymbol{0}}=\text{diag}(\delta_{1n},\delta_{2n},\ldots) \,,
\end{align}
\end{subequations}
are the current operator for the edge $e$ (with the property $\boldsymbol{1}^\intercal \mathbb{J}_e \boldsymbol{\pi}=j_e$) and the projector operator on the state $n$ (with the property $\boldsymbol{1}^\intercal \mathbb{1}^{(n)} \boldsymbol{\pi}=\pi_n$), respectively.

It will also be useful to consider covariances of state observables and currents,
\begin{subequations}
\begin{align} \label{eq:covmat-def-state}
C_{mn}^\mathfrak{s} &\equiv \lim_{t \rightarrow \infty} t^{-1} \langle \theta_m(t) \theta_n(t) \rangle \,, \\ \label{eq:covmat-def-cur}
C_{ee'}^\mathfrak{j} &\equiv \lim_{t \rightarrow \infty} t^{-1} \langle  [\Delta k_{+e}(t)-\Delta k_{-e}(t)] [\Delta k_{+e'}(t)-\Delta k_{-e'}(t)] \rangle \,,
\end{align}
where the superscripts $\mathfrak{s}$ and $\mathfrak{j}$ correspond to covariances of state and current observables, respectively.
\end{subequations}
They can be calculated as~\cite{landi2023current,lapolla2018unfolding,lapolla2019manifestations,lapolla2020spectral}
\begin{subequations} \label{eq:covmat-expression-statecur}
\begin{align} \label{eq:covmat-expression-state}
C_{mn}^\mathfrak{s}&=-\boldsymbol{1}^\intercal \mathds{1}^{(m)} \mathbb{W}^D \mathds{1}^{(n)} \boldsymbol{\pi}-\boldsymbol{1}^\intercal \mathds{1}^{(n)} \mathbb{W}^D \mathds{1}^{(m)} \boldsymbol{\pi} \,, \\ \label{eq:covmat-expression-cur}
C_{ee'}^\mathfrak{j} &=\delta_{ee'} \tau_e -\boldsymbol{1}^\intercal \mathbb{J}_e \mathbb{W}^D \mathbb{J}_{e'} \boldsymbol{\pi}-\boldsymbol{1}^\intercal \mathbb{J}_{e'} \mathbb{W}^D \mathbb{J}_{e} \boldsymbol{\pi} \,,
\end{align}
\end{subequations}
where $\tau_e \equiv W_{+e} \pi_{s(+e)}+W_{-e} \pi_{t(+e)}$ is the undirected traffic at the edge $e$.

\subsection{Static responses}
We also consider the static responses of the state and current observables, that is, the linear response of their steady-state value to some parameter $p$ (e.g., $B_e$ or $S_e$) that controls the transition rates $W_{\pm e}$~\cite{aslyamov2024general}:
\begin{subequations} \label{eq:statrespdef}
\begin{align} 
d_p \mathcal{O} &\equiv \lim_{dp \rightarrow 0} \frac{\mathcal{O}(p+dp)-\mathcal{O}(p)}{dp} \,, \\ d_p \mathcal{J} &\equiv \lim_{dp \rightarrow 0}\frac{\mathcal{J}(p+dp)-\mathcal{J}(p)}{dp}  \,.
\end{align}
\end{subequations}
Operationally, this involves measuring the responses after a time interval following the perturbation of the parameter $p$ that is long enough for the system to relax to its new stationary state. Throughout our paper, we focus on a situation in which the vectors $\boldsymbol{o}$ and $\boldsymbol{x}$ defining the observables do not depend on the perturbed parameter $p$. For such a case, we have
\begin{align}
d_{p} \mathcal{O}=\boldsymbol{o}^\intercal d_{p} \boldsymbol{\pi} \,,
\end{align}
where $d_{p} \boldsymbol{\pi}$ is the static response of the stationary probability vector to the perturbation of transition rates. It can be calculated as~\cite{ptaszynski2024critical}
\begin{align} \label{eq:response-drazin}
d_{p} \boldsymbol{\pi}=-\mathbb{W}^D (d_{p} \mathbb{W}) \boldsymbol{\pi} \,.
\end{align}
We notice that the Drazin inverse form of \cref{eq:response-drazin} is an alternative to the method from Refs.~\cite{aslyamov2024nonequilibrium,aslyamov2024general}. Analogously, the static current responses can be calculated as
\begin{align}
d_p \mathcal{J} = \boldsymbol{x}^\intercal d_p \boldsymbol{j} \,,
\end{align}
where $d_p \boldsymbol{j}=(\ldots,d_p j_e,\ldots)^\intercal$. The responses of edge currents can be calculated as
\begin{align}
d_p j_e=d_p (\boldsymbol{1}^\intercal \mathbb{J}_e \boldsymbol{\pi})=\boldsymbol{1}^\intercal (d_p \mathbb{J}_e) \boldsymbol{\pi}+\boldsymbol{1}^\intercal \mathbb{J}_e d_p \boldsymbol{\pi} \,.
\end{align}
Explicitly, this yields
\begin{subequations} \label{eq:current-expansion}
\begin{align}
d_{B_{e'}} j_e &=\delta_{ee'} j_e+W_{+e} d_{B_{e'}} \pi_{s(+e)}-W_{-e}  d_{B_{e'}} \pi_{t(+e)} \,, \\ \label{eq:current-expansion-asym}
d_{S_{e'}} j_e &=\delta_{ee'} \tau_e/2+W_{+e} d_{S_{e'}} \pi_{s(+e)}-W_{-e}  d_{S_{e'}} \pi_{t(+e)}\,.
\end{align}
\end{subequations}
We note that in the case where the vectors $\boldsymbol{o}$ or $\boldsymbol{x}$ depend on the perturbed parameter $p$, the results presented later can be applied upon replacement $d_p \mathcal{O} \rightarrow d_p \mathcal{O}-\boldsymbol{\pi}^\intercal d_p \boldsymbol{o}$ and $d_p \mathcal{J} \rightarrow d_p \mathcal{J}-\boldsymbol{j}^\intercal d_p \boldsymbol{x}$.

\section{Fluctuation-response relations} \label{sec:frrs}

The main result of our work are the exact identities, called Fluctuation-Response Relations (FRRs), expressing covariances of state and current observables, $\langle\langle\mathcal{O},\mathcal{J}\rangle\rangle$, in terms of the combination of static responses of these observables. They read
\begin{subequations} \label{eq:covar-exact}
\begin{align} 
\label{eq:covar-exact-sym}
\langle \!\langle\mathcal{O},\mathcal{J}\rangle \!\rangle  &= \sum_{e}\frac{\tau_e}{j_e^2}d_{B_e}\mathcal{O}d_{B_e}\mathcal{J}\,, \\
\label{eq:covar-exact-antisym}
 &= \sum_{e} \frac{4}{\tau_e} d_{S_e}\mathcal{O}d_{S_e}\mathcal{J}\,,
\end{align}
\end{subequations}
where, recall, $j_e\equiv W_{+e} \pi_{s(+e)}-W_{-e} \pi_{t(+e)}$ is the directed current at the edge $e$, $\tau_e \equiv W_{+e} \pi_{s(+e)}+W_{-e} \pi_{t(+e)}$ is the undirected traffic, and the parameters $B_e$ and $S_e$ are defined by Eq.~\eqref{eq:rates-model}. These relations complement analogous FRRs for covariances of two currents, $\langle \langle \mathcal{J},\mathcal{J}' \rangle \rangle$, or two state observables, $\langle \langle \mathcal{O},\mathcal{O}' \rangle \rangle$, derived in Refs.~\cite{aslyamov2024frr,ptaszynski2024nonequilibrium}.\footnote{We note that Eq.~\eqref{eq:covar-exact-sym} may appear not applicable when some of the edge currents $j_e$ present in the denominators are equal to zero, in particular at equilibrium. However, when considering a small perturbation of the transition rates $W_{\pm e} \rightarrow W_{\pm e}+\varepsilon W'_{\pm e}$ that makes the edge currents finite, in the limit of $\varepsilon \rightarrow 0$, the denominators, $j_e^2$, and the numerators, $\tau_e d_{B_e}\mathcal{O}d_{B_e}\mathcal{J}$, tend to zero, while their ratios remain finite.} 

The identities~\eqref{eq:covar-exact} result from analogous relations for individual covariance matrix elements,
\begin{subequations}
\label{eq:covmat-exact}
\begin{align} 
\label{eq:covmat-exact-sym}
C^\mathfrak{m}_{ne} &= \sum_{e'}\frac{\tau_{e'}}{j_{e'}^2}d_{B_{e'}} \pi_n d_{B_{e'}} j_e \,, \\ \label{eq:covmat-exact-antisym}
 &= \sum_{e'} \frac{4}{\tau_{e'}} d_{S_{e'}} \pi_n d_{S_{e'}} j_e \,,
\end{align}
\end{subequations}
whose derivation is discussed in the next section.

\subsection{Inverse Fluctuation-Response Relations} \label{subsec:invfrr}

\subsubsection{Inverse FRRs for state responses}
The FRRs~\eqref{eq:covar-exact}--\eqref{eq:covmat-exact} express covariances in terms of the combination of static responses. They can be derived by showing their equivalence to inverse identities, called \textit{Inverse FRRs}, expressing individual state responses $d_p \pi_n$ in terms of the combination of covariances. To show that equivalence, we use Eq.~\eqref{eq:current-expansion} to expand Eq.~\eqref{eq:covmat-exact-sym} as
\begin{align} \nonumber
C^\mathfrak{m}_{ne} &= W_{+e} \sum_{e'}\frac{\tau_{e'}}{j_{e'}^2}d_{B_{e'}} \pi_n d_{B_{e'}} \pi_{s(+e)} \\ \label{eq:covmat-expansion}
&-W_{-e} \sum_{e'}\frac{\tau_{e'}}{j_{e'}^2}d_{B_{e'}} \pi_n d_{B_{e'}} \pi_{t(+e)}+\frac{\tau_e}{j_e} d_{B_e} \pi_n \,.
\end{align}
Equation~\eqref{eq:covmat-exact-antisym} can be expanded analogously. We then apply FRRs for covariances of state observables,
\begin{align} \label{eq:covmat-exact-state}
C^\mathfrak{s}_{mn} = \sum_{e}\frac{\tau_e}{j_e^2}d_{B_e} \pi_m d_{B_e} \pi_n = \sum_{e} \frac{4}{\tau_e} d_{S_e} \pi_m d_{S_e} \pi_n\,,
\end{align}
that have been derived in the companion Letter~\cite{ptaszynski2024nonequilibrium}. 
Using them, we identify sums in \cref{eq:covmat-expansion} with the covariances $C^\mathfrak{s}_{ns(+e)}$ and $C^\mathfrak{s}_{nt(+e)}$. As a result, we obtain the mentioned Inverse FRRs for state responses,
\begin{align} \label{eq:inverse-frr}
\frac{\tau_e}{j_e} d_{B_e} \pi_n =
2d_{S_e} \pi_n =C^\mathfrak{m}_{ne}-W_{+e} C^\mathfrak{s}_{ns(+e)}+W_{-e} C^\mathfrak{s}_{nt(+e)} \,.
\end{align}
These identities are derived in Appendix~\ref{app:proof-frr} using an algebraic approach similar to that used in the companion Letter~\cite{ptaszynski2024nonequilibrium} for deriving \cref{eq:covmat-exact-state}. In Appendix~\ref{app:proof-invfrr-zheng}, we also present an alternative derivation based on the recent framework from Ref.~\cite{zheng2025spatial}, which can be regarded as a development of the nonequilibrium FDTs from Refs.~\cite{agarwal1972fluctuation,baiesi2009fluctuations,seifert2010fluctuation}.

\subsubsection{Inverse FRRs for current responses}
A certain form of analogous Inverse FRRs, expressing the current responses $d_p j_e$ in terms of covariances, can now be obtained by inserting \cref{eq:inverse-frr} into Eq.~\eqref{eq:current-expansion}. However, using explicit derivation (see Appendices~\ref{app:proof-frr}--\ref{app:proof-invfrr-zheng}) we can obtain more elegant expressions
\begin{align} \label{eq:inverse-frr-cur}
\frac{\tau_{e'}}{j_{e'}} d_{B_{e'}} j_e =
2 d_{S_{e'}} j_e =C^\mathfrak{j}_{ee'}-W_{+e'} C^\mathfrak{m}_{s(+e')e}+W_{-e'} C^\mathfrak{m}_{t(+e')e} \,.
\end{align}
We now recall that at equilibrium, the mixed covariances $C^\mathfrak{m}_{ne}$ vanish due to the time-reversal symmetry~\cite{dieball2022correlations}. Consequently, \cref{eq:inverse-frr-cur} leads to the well-known fluctuation-dissipation theorem $d_{S_{e'}} j_e =C^\mathfrak{j}_{ee'}/2$. 

The other outcome of Eq.~\eqref{eq:inverse-frr-cur} is its relation to nonreciprocity, i.e., breaking of the Onsager symmetry $d_{S_{e'}} j_e=d_{S_e} j_{e'}$ that can occur far from equilibrium. It can be quantified using the nonreciprocity measure $\mathcal{N}_{ee'} \equiv d_{S_{e'}} j_e-d_{S_e} j_{e'}$. Using Eq.~\eqref{eq:inverse-frr-cur} and the symmetry of covariance matrix elements $C^\mathfrak{j}_{ee'}=C^\mathfrak{j}_{e'e}$, we found this measure to be strictly related to state-current covariances,
\begin{align} \label{eq:nonrec}
&\mathcal{N}_{ee'} \\ \nonumber &=\frac{1}{2} \left(W_{+e} C^\mathfrak{m}_{s(+e)e'}-W_{-e} C^\mathfrak{m}_{t(+e)e'}-W_{+e'} C^\mathfrak{m}_{s(+e')e}+W_{-e'} C^\mathfrak{m}_{t(+e')e} \right) \,.
\end{align}
Consequently, the breaking of the Onsager symmetry requires the presence of correlations of the state and current observables. 

Finally, an additional constraint on the state-current covariances holds in a situation where the system is out of equilibrium but two edge currents vanish, $j_e=j_{e'}=0$. Then, as proven in Ref.~\cite{altaner2016fluctuation}, one obtains the relation $C^\mathfrak{j}_{ee'}=d_{S_{e'}} j_e+d_{S_e} j_{e'}$ that generalizes the equilibrium fluctuation-dissipation theorem. Using Eq.~\eqref{eq:inverse-frr-cur}, this implies
\begin{align}
W_{+e} C^\mathfrak{m}_{s(+e)e'}-W_{-e} C^\mathfrak{m}_{t(+e)e'}+W_{+e'} C^\mathfrak{m}_{s(+e')e}-W_{-e'} C^\mathfrak{m}_{t(+e')e}=0 \,.
\end{align}
We emphasize that, away from equilibrium, the individual covariances included in the expression above may be nonvanishing. Adding or substracting 1/2 of the left-hand side of the above expression to Eq.~\eqref{eq:nonrec}, the nonreciprocity parameter simplifies to
\begin{align} \nonumber
\mathcal{N}_{ee'} &=W_{+e} C^\mathfrak{m}_{s(+e)e'}-W_{-e} C^\mathfrak{m}_{t(+e)e'} \\ &=W_{-e'} C^\mathfrak{m}_{t(+e')e}-W_{+e'} C^\mathfrak{m}_{s(+e')e} \,.
\end{align}

\subsection{Generalization to generic flux observables} \label{subsec:fluxobs}
In our paper, we focus on correlations between state observables and time-antisymmetric current observables [Eqs.~\eqref{eq:stateobsdef}--\eqref{eq:curobsdef}], since such observables are most typically considered in the literature. Here we generalize our framework to correlations between state observables and generic flux observables (without the time-antisymmetry property) defined as
\begin{align}
\hat{F}(t) \equiv \frac{1}{t} \sum_{\pm e} y_{\pm e} k_{\pm e}(t)\,,
\end{align}
where $\sum_{\pm e}$ denotes the summation over the forward and backward directed edges, and $k_{\pm e}(t)$ has been defined below Eq.~\eqref{eq:curobsdef}. Observables of this type attracted some attention, e.g., in the context of bounds on entropy production~\cite{bakewell2023general,pietzonka2024thermodynamic}. A well-known example is the total traffic $\hat{T}(t)=t^{-1}\sum_{\pm e} k_{\pm e}(t)$ that measures the total number of jumps in the Markov network in the time-interval $[0,t]$. The average flux observables can be expressed as
\begin{align} \label{eq:fluxobsdef}
\mathcal{F} \equiv \lim_{t \rightarrow \infty} \langle \hat{F}(t) \rangle=\sum_{\pm e} y_{\pm e} \varphi_{\pm e} \,,
\end{align}
where $\varphi_{\pm e} \equiv W_{\pm e} \pi_{s(\pm e)}$ is the unidirectional probability flux along the edge $\pm e$.

We now observe that FRRs~\eqref{eq:covar-exact} do not hold if we simply replace $\mathcal{J}$ by $\mathcal{F}$. This is because their derivation relies on the identity $(\tau_{e'}/j_{e'}) d_{B_{e'}} j_e=2d_{S_{e'}} j_e$~\cite{aslyamov2024nonequilibrium} (see Appendix~\ref{app:proof-frr}) and \cref{eq:covmat-expansion}, which do not apply to unidirectional fluxes $\varphi_{\pm e}$. However, covariances of state and flux observables obey a different FRR
\begin{align} \label{eq:frr-gen}
\langle\! \langle \mathcal{O},\mathcal{F} \rangle \! \rangle \equiv \lim_{t \rightarrow \infty} t \langle \Delta \hat{o}(t) \Delta \hat{F}(t) \rangle=\sum_{\pm e} \frac{1}{\varphi_{\pm e}} d_{X_{\pm e}}\mathcal{O}d_{X_{\pm e}}\mathcal{F} \,,
\end{align} where $X_{\pm e} \equiv \ln W_{\pm e}$. This FRR is derived in Appendix~\ref{app:der-frr-gen}. Its main difference compared to FRRs in Eq.~\eqref{eq:covar-exact} is that it is expressed in terms of responses to independent perturbations of transition rates $W_{\pm e}$, while the former are expressed in terms of responses to symmetric and antisymmetric edge parameters $B_e$ and $S_e$, which affect both forward and backward transition rates $W_{+e}$ and $W_{-e}$ [see Eq.~\eqref{eq:rates-model}].

\section{Application for calculating fluctuations} \label{sec:calc-fluct}

\subsection{Unicyclic networks}
\begin{figure}
    \centering
    \includegraphics[width=0.7\linewidth]{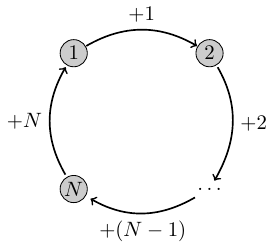}
    \caption{Scheme of a unicyclic network with each state $n \in \{1,N-1\}$ being a source of a single edge $+n$ pointing to the state $n+1$, and the edge $+N$ pointing from $N$ to $1$. The transtion $+N$ is considered unidirectional ($W_{-N}=0$).}
    \label{fig:fig-unicylic}
\end{figure}

We now show that FRRs~\eqref{eq:covar-exact}--\eqref{eq:covmat-exact} can be used to analytically calculate fluctuations in Markov networks when the stationary state $\boldsymbol{\pi}$ can be analytically determined. Importantly, this method does not involve the Drazin inverse present in Eqs.~\eqref{eq:covmat-expression} and~\eqref{eq:covmat-expression-statecur}, which for large systems is usually analytically intractable. We note that the fluctuations can be also calculated analytically (even without directly determining $\boldsymbol{\pi}$) using the characteristic polynomial method~\cite{wachtel2015fluctuating,bruderer2014inverse}. However, our method is still advantageous in providing an explicit decomposition of covariances into contributions associated with responses to perturbations of different edges. As will be illustrated in Sec.~\ref{sec:examples}, this helps one understand the properties of fluctuations (e.g., their sign) based on response properties of the Markov network.

We first demonstrate our idea on the example of unicyclic networks with at least one unidirectional transition, presented in Fig.~\ref{fig:fig-unicylic}. Such networks describe various physically relevant setups, e.g., enzymatic reactions~\cite{barato2015skewness} or electronic transport~\cite{gustavsson2006counting,gustavsson2009electron}. A simple example of such a network is Michaelis-Menten reaction scheme
\begin{align}
\ce{E + S <=>[W_{+1}][W_{-1}] ES ->[W_{+2}] P + E} \,,
\end{align}
where the enzyme E switches between unbound state (state $1$) and the enzyme-substrate complex ES (state $2$), and the unidirectional transition $+2$ corresponds to the release of the product P.

The steady state of such networks can be determined analytically. To that end, we note that due to Kirchhoff's law all edge currents are equal, $\forall_e \, j_e=j$, with the current equal to
\begin{align}
j=W_{+N} \pi_N \,,
\end{align}
so that $\pi_N=j/W_{+N}$. Using the formula for the edge current, $j=W_{+n} \pi_n-W_{-n} \pi_{n+1}$, the probabilities of states $\pi_n$ with $n<N$ can be then determined iteratively as
\begin{align}
\pi_{n}=\frac{j+W_{-n} \pi_{n+1}}{W_{+n}} \,.
\end{align}
The current $j$ can finally be determined using the normalization condition $\sum_{n=1}^N \pi_n=1$. Using analytic formulas for the state probabilities and the current $j$, the covariances $C^\mathfrak{m}_{ne}$ can be calculated using Eq.~\eqref{eq:covmat-exact}. We note that for unidirectional transitions, the parametrization~\eqref{eq:rates-model} may appear to be ill-defined. However, the responses to symmetric or antisymmetric perturbations are well-defined and given by the formulas~\cite{owen2020universal}
\begin{subequations}
\begin{align}
d_{B_n} &=W_{+n} d_{W_{+n}}+W_{-n} d_{W_{-n}} \,, \\
d_{S_n} &=\left( W_{+n} d_{W_{+n}}-W_{-n} d_{W_{-n}} \right)/2 \,,
\end{align}
\end{subequations}
where, recall, $d_p$ is our shortened notation for the static response [see \cref{eq:statrespdef}]. This expression can be applied to unidirectional transitions with $W_{-n}=0$ by taking $W_{-n} d_{W_{-n}}=0$. In networks with only unidirectional transitions ($\forall_n \, W_{-n}=0)$ the state probabilities, current, and their responses, are given by the explicit formulas
\begin{subequations}
\begin{align}
\pi_n&=j/W_{+n} \,, \\
j&=\left( \sum_{n=1}^N W_{+n}^{-1} \right)^{-1} \,, \\
d_{B_n} \pi_m &= 2 d_{S_n} \pi_m =\pi_n(\pi_m-\delta_{nm}) \,, \\
d_{B_n}j &= 2d_{S_n}j = \pi_n j \,.
\end{align}
\end{subequations}
Using Eq.~\eqref{eq:covmat-exact} with $\forall_e: \, j_e=\tau_e=j$, the mixed covariances can be then calculated as
\begin{align}
\forall_e: \;  C^\mathfrak{m}_{ne}= -\pi_n \left( \pi_n-\sum_{k=1}^N \pi_k^2 \right)=-\frac{j^2}{W_{+n}^2}+ \frac{j^3}{W_{+n}} \sum_{k=1}^N \frac{1}{W_{+k}^{2}} \,.
\end{align}

\subsection{Birth-and-death processes}
\begin{figure}
    \centering
    \includegraphics[width=\linewidth]{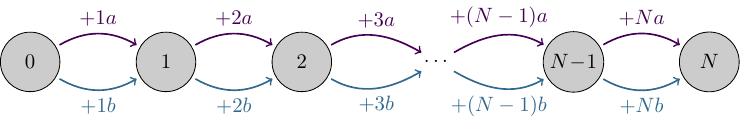}
    \caption{Scheme of one-dimensional Markov network with each state $n \in \{0,N\}$ apart from $n=N$ being a source of several edges $+(n+1)\nu$ (here $\nu \in \{a,b\}$) pointing to a tip $n+1$. The index $\nu$ denotes the channel of transition.}
    \label{fig:fig-one-dim}
\end{figure}

As a second example, let us consider one-dimensional Markov networks (so-called birth-and-death processes~\cite{ledermann1954spectral}) presented in Fig.~\ref{fig:fig-one-dim}. Such models have been applied in many contexts, e.g., to describe chemical bistability (Schl{\"o}gl model)~\cite{schlogl1972chemical,vellela2009stochastic,ge2009thermodynamic}, bistable electric circuits~\cite{landauer1962fluctuations,hanggi1982stochastic}, magnetic systems (Curie-Weiss model)~\cite{herpich2020njp,meibohm2022finite,meibohm2023landau,ptaszynski2024critical}, coupled heat engines~\cite{vroylandt2018collective,vroylandt2020efficiency}, or disease spread~\cite{nieddu2022characterizing}. We note that models with either finite $N$ (such as Curie-Weiss model) or $N \rightarrow \infty$ (such as Schl{\"o}gl model) can be considered within the same framework. Different transition channels $\nu$ denoted in Fig.~\ref{fig:fig-one-dim} may correspond, e.g., to transitions induced by different reservoirs. To illustrate that on the example, let us consider the Schl{\"o}gl model in which two channels $\nu \in \{a,b \}$ correspond to different chemical reactions,
\begin{subequations}
\begin{align}
\text{channel $a$}: \quad &\ce{A + 2X <=>[k_{+a}][k_{-a}] 3X} \,, \\
\text{channel $b$}: \quad &\ce{B <=>[k_{+b}][k_{-b}] X} \,,
\end{align}
\end{subequations}
where $k_{\pm \nu}$ are reaction constants. The transition rates in the model can be expressed as~\cite{vellela2009stochastic}
\begin{subequations}
\begin{align}
W_{+na} &=\frac{c_A k_{+a} (n-1)(n-2)}{\Omega} \,, \\
W_{-na} &=\frac{k_{-a} n(n-1)(n-2)}{\Omega^2} \,, \\
W_{+nb} &=c_B k_{+b} \Omega \,, \\
W_{-nb} &=n k_{-b} \,,
\end{align}
\end{subequations}
where $n$ is the number of molecules $X$, $c_A$ and $c_B$ are concentrations of species $A$ and $B$ that are kept constant (chemostated), and $\Omega$ is the volume.

For the class of systems considered, although transitions may occur in several channels $\nu$ (which can drive the system out of equilibrium), the net probability currents between the system states $\pi_n$ vanish (i.e., the system is detailed balanced). As a result, the stationary state $\boldsymbol{\pi}$ can be determined analytically. To that end one defines the total transition rate
\begin{align}
W_{\pm n} = \sum_\nu W_{\pm n \nu} \,.
\end{align}
The steady state can be then determined as
\begin{align}
\pi_n=\pi_0 \prod_{m=1}^n \frac{W_{+m}}{W_{-m}} \,,
\end{align}
with $\pi_0$ given by the normalization condition $\sum_{n=0}^N \pi_n=1$. Consequently, the covariance of occupation of state $m$ with the current through the edge $n\nu$ can be determined analytically using \cref{eq:covar-exact} as
\begin{subequations}
\begin{align} 
C^\mathfrak{m}_{m,n\nu} &= \sum_{k=1}^N \sum_{\nu'} \frac{\tau_{k\nu'}}{j_{k\nu'}^2} d_{B_{k\nu'}} \pi_m d_{B_{k\nu'}} j_{n\nu}  \\
 &= \sum_{k=1}^N \sum_{\nu'} \frac{4}{\tau_{k\nu'}} d_{S_{k\nu'}} \pi_m d_{S_{k\nu'}} j_{n\nu} \,,
\end{align}
\end{subequations}
with
\begin{subequations}
\begin{align}
d_{B_{n \nu}} &= W_{+n\nu} d_{W_{+n\nu}}+W_{-n\nu} d_{W_{-n\nu}} \,, \\
d_{S_{n \nu}} &= \left(W_{+n\nu} d_{W_{+n\nu}}-W_{-n\nu} d_{W_{-n\nu}} \right)/2\,, \\
j_{n\nu} &= W_{+n \nu} \pi_{n-1}-W_{-n \nu} \pi_n \,, \\
\tau_{n\nu} &= W_{+n \nu} \pi_{n-1}+W_{-n \nu} \pi_n \,.
\end{align}
\end{subequations}

\section{Examples} \label{sec:examples}

\subsection{Transport through a quantum dot}

\begin{figure}
    \centering
    \includegraphics[width=\linewidth]{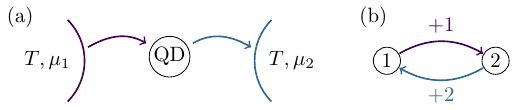}
    \caption{(a) Scheme of a quantum dot connected to two reservoirs $1$ and $2$. The applied voltage induces transitions between states occupied by $N$ and $N+1$ electrons. The voltage is much larger than the thermal energy $k_B T$, so that the electrons transitions can be considered unidirectional. (b) A Markov network describing the system. Here, labels $\{1,2\}$ denote the states with $N$ and $N+1$ electrons.}
    \label{fig:fig-QD}
\end{figure}

Let us now present how our results can be used to gain insight into the behavior of fluctuations in some physically relevant systems. As a first example, let us consider a single quantum dot model presented in Fig.~\ref{fig:fig-QD}. This model describes the experimental setup from Refs.~\cite{gustavsson2006counting,gustavsson2009electron}. In those experiments, charge states and electron transitions were monitored in real time, which could enable experimental verification of our results. For the model considered, the covariance between the occupation of the state $1$ and the particle current $j=j_1=j_2$ can be determined as
\begin{align} \label{eq:qd-covar-1}
C^\mathfrak{m}_{11}=C^{\mathfrak{m}(1)}_{11}+C^{\mathfrak{m}(2)}_{11} \,,
\end{align}
where
\begin{subequations}
\begin{align}
C^{\mathfrak{m}(1)}_{11} &\equiv -\frac{|d_{B_1} \pi_1 d_{B_1} j|}{j}=-\pi_1^2(1-\pi_1) \,, \\ 
C^{\mathfrak{m}(2)}_{11} &\equiv \frac{|d_{B_2} \pi_1 d_{B_2} j|}{j}=\pi_2 \pi_1\,,
\end{align}
\end{subequations}
$\pi_1=W_{+2}/(W_{+1}+W_{+2})$, and $\pi_2=1-\pi_1$. We note that the first (second) term is negative (positive) due to $d_{B_1} \pi_1 d_{B_1} j<0$ ($d_{B_2} \pi_1 d_{B_2} j>0$). In fact, the enhancement of the transition rate $W_{+1}$ increases the current and reduces the population of the state $\pi_1$, while the enhancement of the transition $W_{+2}$ increases both the current and the population $\pi_1$. 

\begin{figure}
    \centering
    \includegraphics[width=0.95\linewidth]{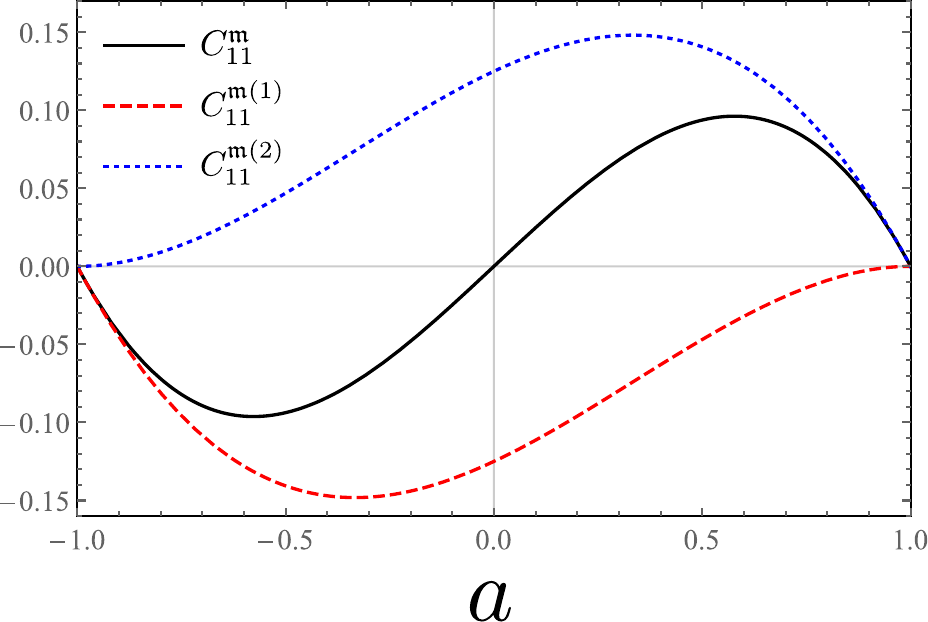}
    \caption{The covariance $C^\mathfrak{m}_{11}$ and its individual components $C_{11}^{\mathfrak{m}(e)}$ as a function of the asymmetry coefficient $a$.}
    \label{fig:covar-qd}
\end{figure}
The covariance $C^\mathfrak{m}_{11}$ can be further expressed in a simple form
\begin{align} \label{eq:covar-qd-fun-a}
C^\mathfrak{m}_{11}=\frac{a}{4} (1-a^2) \,,
\end{align}
where $a=(W_{+1}-W_{+2})/(W_{+1}+W_{+2})$ is the asymmetry coefficient. The covariance $C^\mathfrak{m}_{11}$ and its individual components $C^{\mathfrak{m}(e)}_{11}$ are plotted as a function of the coefficient $a$ in Fig.~\ref{fig:covar-qd}. Based on this plot, we now unravel the physical meaning of Eq.~\eqref{eq:covar-qd-fun-a}. We note that for the unidirectional transitions considered, the steady state and the stationary current are determined mainly by the slowest process. Consequently, the response to the perturbation of a smaller transition rate is larger. Thus, when $W_{+1}<W_{+2}$ ($a<0$), the term $C^{\mathfrak{m}(1)}_{11}$ dominates and the covariance $C^\mathfrak{m}_{11}$ becomes negative. In contrast, in the opposite regime of $W_{+1}>W_{+2}$ ($a>0$), the term $C^{\mathfrak{m}(2)}_{11}$ dominates and the covariance becomes positive. This shows that the covariance of the state and current observables provide information about the sign of the asymmetry coefficient $a$. In contrast, such information is not provided by fluctuations of state and current observables alone, which only depend on the absolute value of $a$:
\begin{subequations}
\begin{align}
C_{11}^\mathfrak{s}=C_{22}^\mathfrak{s}=-C_{12}^\mathfrak{s}&=\frac{1}{4\Gamma}(1-a^2) \,, \\ C_{11}^\mathfrak{j}=C_{22}^\mathfrak{j}=C_{12}^\mathfrak{j}&=\frac{\Gamma}{4}(1-a^4) \,,
\end{align}
\end{subequations}
where $\Gamma=(W_{+1}+W_{+2})/2$.

\subsection{Enzymatic inhibition}
\begin{figure}
    \centering
    \includegraphics[width=0.7\linewidth]{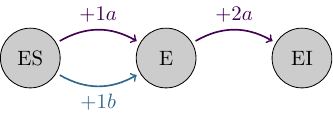}
    \caption{Scheme of a Markov model corresponding to reaction~\eqref{eq:inh}. The system can reside in three states $n \in \{\text{ES},\text{E},\text{EI} \}$, corresponding to enzyme-substrate complex, unbound enzyme, and enzyme-inhibitor complex. The transition $+1b$ is associated with the release of the product P.}
    \label{fig:fig-enz}
\end{figure}

We now consider a more complex model, corresponding to the competitive enzymatic inhibition scheme,
\begin{align} \label{eq:inh}
\ce{EI + S <=>[W_{-2a}][W_{+2a}] E + S + I <=>[W_{-1a}][W_{+1a}] ES + I ->[W_{+1b}] P + I + E} \,.
\end{align}
The corresponding Markov network is presented in Fig.~\ref{fig:fig-enz}.  We focus on the covariance between the time spent in the unbound state E and the rate of product release $j_{1b}$. This quantity is experimentally relevant, as occupations of bound and unbound states~\cite{lerner2018toward}, as well as individual product release events~\cite{english2006ever}, can be monitored in real-time at the single molecule level. It can be calculated as
\begin{align}
C^\mathfrak{m}_{\text{E},1b}=C_{\text{E},1b}^{\mathfrak{m}(1a)}+C_{\text{E},1b}^{\mathfrak{m}(1b)}+C_{\text{E},1b}^{\mathfrak{m}(2a)} \,,
\end{align}
where
\begin{align} \label{eq:ke1b}
C_{\text{E},1b}^{\mathfrak{m}(e)} \equiv \frac{4}{\tau_e} d_{S_e} \pi_\text{E} d_{S_e} j_{1b} \,.
\end{align}
The explicit expressions for  $C_{\text{E},1b}^{\mathfrak{m}(e)}$ are presented in the Appendix~\ref{app:exp-expr-ke1b}. Interestingly, we find
\begin{figure}
    \centering
    \includegraphics[width=0.95\linewidth]{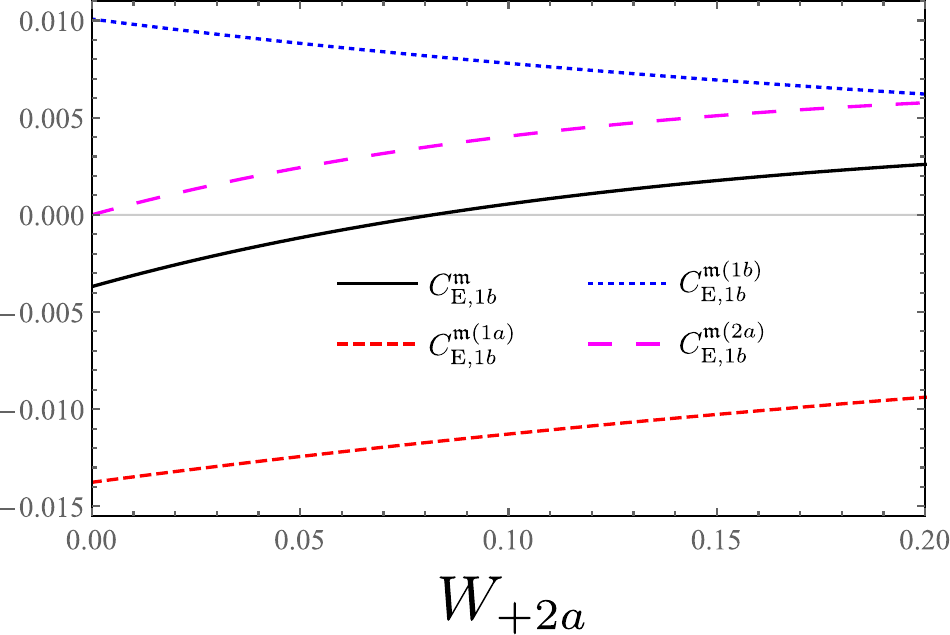}
    \caption{The covariance $C^\mathfrak{m}_{\text{E},1b}$ and its individual components $C_{\text{E},1b}^{\mathfrak{m}(e)}$ as a function of the transition rate $W_{+2a}$ for $W_{+1a}=1$, $W_{-1a}=W_{-2a}=0.5$, $W_{+1b}=0.05$.}
    \label{fig:covar-enz}
\end{figure}
\begin{align}
C_{\text{E},1b}^{\mathfrak{m}(1a)}<0, \quad C_{\text{E},1b}^{\mathfrak{m}(1b)} \,,\, C_{\text{E},1b}^{\mathfrak{m}(2a)}>0 \,.
\end{align}
Inequality $C_{\text{E},1b}^{\mathfrak{m}(1a)}<0$ results from the fact that by enhancing the transition $+1a$, one increases the probability of state E ($d_{S_{1a}} \pi_\text{E}>0$) while reducing the rate of product release ($d_{S_{1a}} j_{1b}<0$) due to decrease of the population of the enzyme-substrate complex ES. On the other hand, by enhancing transition $+1b$, one increases both the rate of product release ($d_{S_{1b}} j_{1b}>0$) and the population of state E ($d_{S_{1b}} \pi_\text{E}>0$), and thus $C_{\text{E},1b}^{\mathfrak{m}(1b)}>0$. Finally, by enhancing transition $+2a$ one reduces both the population of state E ($d_{S_{2a}} \pi_\text{E}<0$) and the rate of product release ($d_{S_{2a}} j_{1b}<0$) due to the inhibition effect, i.e., trapping in state EI. Consequently, $C_{\text{E},1b}^{\mathfrak{m}(2a)}>0$. 

In Fig.~\ref{fig:covar-enz}, we plot the behavior of covariance $C^\mathfrak{m}_{\text{E},1b}$ and its individual components $C_{\text{E},1b}^{\mathfrak{m}(e)}$ as a function of the transition rate $W_{+2a}$ that is proportional to the inhibitor concentration. As shown, for the parameters considered, the contribution $C_{\text{E},1b}^{(1a)}$ dominates for small $W_{+2a}$, leading to negative covariance. By increasing $W_{+2a}$, one magnifies the contribution $C_{\text{E},1b}^{\mathfrak{m}(2a)}$ related to inhibition effects, making the covariance $C_{\text{E},1b}^{\mathfrak{m}}$ positive. This shows that the sign of $C^\mathfrak{m}_{\text{E},1b}$ can provide information about the magnitude of inhibition effects.

\section{Concluding remarks} \label{sec:concl}

We note that analogously to FRRs for state observables~\cite{ptaszynski2024nonequilibrium} and currents~\cite{aslyamov2024frr}, FRRs~\eqref{eq:covar-exact} have an intuitive interpretation: They mean that the state and current observables can be positively (negatively) correlated only when they respond with the same (opposite) sign to at least one symmetric and antisymmetric edge perturbation. Beyond providing a fundamental link between fluctuations and response, this result has practical relevance putting constraints on Markov models of physical systems that are consistent with measured data. An interesting perspective for future research is to extend our result to continuous-space Langevin dynamics~\cite{gao2022thermodynamic,gao2024thermodynamic}, where covariances of state and current observables have recently received a certain interest~\cite{dieball2022correlations}.

\begin{acknowledgments}
K.P., T.A. and M.E. acknowledge the financial support from, respectively, project No.\ 2023/51/D/ST3/01203 funded by the National Science Centre, Poland, 
project ThermoElectroChem (C23/MS/18060819) from Fonds National de la Recherche-FNR, Luxembourg,  
project TheCirco (INTER/FNRS/20/15074473) funded by FRS-FNRS (Belgium) and FNR (Luxembourg).
\end{acknowledgments}

\appendix

\section{Derivation of Eq.~\eqref{eq:covmat-expression}} \label{app:der-covmat-expression}

Using theory from Ref.~\cite{utsumi2007full}, the covariances $C^\mathfrak{m}_{ne}$ can be calculated as
\begin{align}
    C^\mathfrak{m}_{ne} = \frac{\partial}{\partial \zeta_n}\frac{\partial}{\partial q_{e}} \lambda(\boldsymbol{q},\boldsymbol{\zeta})\Big|_{\boldsymbol{q},\boldsymbol{\zeta}=\boldsymbol{0}}\,,
\end{align}
where $\lambda(\boldsymbol{q},\boldsymbol{\zeta})$ is the eigenvalue of the tilted rate matrix $\mathbb{W}^\phi(\boldsymbol{q},\boldsymbol{\zeta})$ with the largest real part;
this eigenvalue $\lambda(\boldsymbol{q},\boldsymbol{\zeta})$ and the corresponding eigenvector $\boldsymbol{v}(\boldsymbol{q},\boldsymbol{\zeta})$ satisfy
\begin{align}
    \label{eq:eigenval-1}
    \mathbb{W}^\phi(\boldsymbol{q},\boldsymbol{\zeta})\boldsymbol{v}(\boldsymbol{q},\boldsymbol{\zeta}) = \lambda(\boldsymbol{q},\boldsymbol{\zeta})\boldsymbol{v}(\boldsymbol{q},\boldsymbol{\zeta})\,,
\end{align}
with $\lambda(\boldsymbol{0},\boldsymbol{0})=0$ and $\boldsymbol{v}(\boldsymbol{0},\boldsymbol{0})=\boldsymbol{\pi}$. Acting on both sides of \cref{eq:eigenval-1} with the derivative $\partial_{\zeta_n}\partial_{q_e}$, we obtain
\begin{align}
    \label{eq:eigenval-4}
    &(\partial_{\zeta_n}\mathbb{W}^\phi)\partial_{q_{e}}\boldsymbol{v} + (\partial_{q_{e}}\mathbb{W}^\phi)\partial_{\zeta_n}\boldsymbol{v}+\mathbb{W}^\phi(\partial_{q_{e}}\partial_{\zeta_n}\boldsymbol{v})  \nonumber\\
    &=(\partial_{\zeta_{n}}\partial_{q_e}\lambda)\boldsymbol{v}+(\partial_{\zeta_n}\lambda)(\partial_{q_e}\boldsymbol{v})+ (\partial_{q_{e}}\lambda)\partial_{\zeta_n}\boldsymbol{v}+\lambda(\partial_{\zeta_n}\partial_{q_e}\boldsymbol{v})\,,
\end{align}
where we used $\partial_{\zeta_n}\partial_{q_e}\mathbb{W}^\phi(\boldsymbol{q},\boldsymbol{\zeta}) = 0$. At $\boldsymbol{q},\boldsymbol{\zeta}=\boldsymbol{0}$, we have
\begin{align}
    \label{eq:eigenval-5}
    &\mathds{1}^{(n)}\partial_{q_{e}}\boldsymbol{v} + \mathbb{J}_e \partial_{\zeta_n}\boldsymbol{v}+\mathbb{W}(\partial_{q_{e}}\partial_{\zeta_n}\boldsymbol{v})  \nonumber\\
    &=(\partial_{\zeta_{n}}\partial_{q_e}\lambda)\boldsymbol{\pi}+(\partial_{\zeta_n}\lambda)(\partial_{q_e}\boldsymbol{v})+ (\partial_{q_{e}}\lambda)\partial_{\zeta_n}\boldsymbol{v}\,,
\end{align}
where we used $\lambda(\boldsymbol{0},\boldsymbol{0})=0$, $\boldsymbol{v}(\boldsymbol{0},{\boldsymbol{0}})=\boldsymbol{\pi}$, $\mathbb{W}^\phi(\boldsymbol{0},\boldsymbol{0})=\mathbb{W}$, and definitions from Eq.~\eqref{eq:curprojop}. We further multiply both sides of \cref{eq:eigenval-5} by $\boldsymbol{1}^\intercal$ and notice $\boldsymbol{1}^{\intercal}\mathbb{W}=\boldsymbol{0}$. Since $\boldsymbol{v}$ is defined up to normalization, we take $\boldsymbol{1}^\intercal \boldsymbol{v}=1$, so that $\boldsymbol{1}^\intercal \partial_{\zeta_n} \boldsymbol{v}=0$ and $\boldsymbol{1}^\intercal \partial_{q_e} \boldsymbol{v}=0$. We obtain
\begin{align} \label{eq:eigenval-5a}
C^\mathfrak{m}_{ne}=\partial_{\zeta_n}\partial_{q_e}\lambda = \boldsymbol{1}^\intercal \mathbb{J}_e \partial_{\zeta_{n}}\boldsymbol{v} + \boldsymbol{1}^\intercal \mathds{1}^{(n)} \partial_{q_e}\boldsymbol{v}\,.
\end{align}
To determine the derivative $\partial_{\zeta_{n}}\boldsymbol{v}$, we apply the derivative $\partial_{\zeta_{n}}$ to both sides of Eq.~\eqref{eq:eigenval-1},
\begin{align}
    \label{eq:eigenval-6}
    (\partial_{\zeta_n} \mathbb{W}^\phi)\boldsymbol{v}+ \mathbb{W}^\phi \partial_{\zeta_n} \boldsymbol{v}= (\partial_{\zeta_n} \lambda )\boldsymbol{v}+\lambda \partial_{\zeta_n} \boldsymbol{v}\,.
\end{align}
At $\boldsymbol{q},\boldsymbol{\zeta}=\boldsymbol{0}$, we have
\begin{align}
    \label{eq:eigenval-7}
     (\partial_{\zeta_n} \mathbb{W}^\phi) \boldsymbol{\pi}+ \mathbb{W} \partial_{\zeta_n} \boldsymbol{v}= (\partial_{\zeta_n} \lambda )\boldsymbol{\pi}\,,
\end{align}
where we again used $\lambda(\boldsymbol{0},\boldsymbol{0})=0$, $\boldsymbol{v}(\boldsymbol{0},{\boldsymbol{0}})=\boldsymbol{\pi}$, $\mathbb{W}^\phi(\boldsymbol{0},\boldsymbol{0})=\mathbb{W}$. We then act on both sides of \cref{eq:eigenval-7} with the Drazin inverse $\mathbb{W}^D$ and use $\mathbb{W}^D \boldsymbol{\pi}=0$ and $\mathbb{W}^D \mathbb{W} \partial_{\zeta_n} \boldsymbol{v}=(\mathds{1}-\boldsymbol{\pi} \boldsymbol{1}^\intercal)\partial_{\zeta_n} \boldsymbol{v} =\partial_{\zeta_n} \boldsymbol{v}$ due to $\boldsymbol{1}^\intercal \partial_{\zeta_n} \boldsymbol{v}=0$. We obtain 
\begin{align}   \label{eq:eigenval-8}
\partial_{\zeta_n} \boldsymbol{v}=-\mathbb{W}^D  (\partial_{\zeta_n} \mathbb{W}^\phi) \boldsymbol{\pi}= -\mathbb{W}^D  \mathds{1}^{(n)} \boldsymbol{\pi}\,,
\end{align}
where we used definitions from \cref{eq:curprojop}. Applying the same procedure for $\partial_{q_e} \boldsymbol{v}$, we get
\begin{align} \label{eq:eigenval-9}
\partial_{q_e} \boldsymbol{v}=-\mathbb{W}^D  (\partial_{q_e} \mathbb{W}^\phi) \boldsymbol{\pi}= -\mathbb{W}^D  \mathbb{J}_e \boldsymbol{\pi}\,.
\end{align}
Inserting Eqs.~\eqref{eq:eigenval-8}--\eqref{eq:eigenval-9} to \cref{eq:eigenval-5a}, we obtain Eq.~\eqref{eq:covmat-expression}.

\section{Algebraic derivation of Inverse FRRs \eqref{eq:inverse-frr}--\eqref{eq:inverse-frr-cur}} \label{app:proof-frr}

Here we derive Inverse FRRs~\eqref{eq:inverse-frr}--\eqref{eq:inverse-frr-cur}, which provide a basis for our main result, FRRs~\eqref{eq:covar-exact}--\eqref{eq:covmat-exact} (see Sec.~\ref{subsec:invfrr}). To that end, we first use Eqs.~\eqref{eq:covmat-expression} and~\eqref{eq:covmat-expression-state} to express right-hand side of Eq.~\eqref{eq:inverse-frr} as
\begin{align} \nonumber
&C^\mathfrak{m}_{ne}-W_{+e} C^\mathfrak{s}_{ns(+e)}+W_{-e} C^\mathfrak{s}_{nt(+e)} \\ \label{eq:rhs-inversefrr-exp}
&=-\boldsymbol{1}^\intercal \mathbb{D}_e \mathbb{W}^D \mathds{1}^{(n)} \boldsymbol{\pi}-\boldsymbol{1}^\intercal \mathds{1}^{(n)} \mathbb{W}^D \mathbb{D}_e \boldsymbol{\pi} \,,
\end{align}
where $\mathbb{D}_e=\mathbb{J}_e-W_{+e}\mathds{1}^{(s(+e))}+W_{-e} \mathds{1}^{(t(+e))}$. This matrix explicitly reads
\begin{align}
\mathbb{D}_e=
\begin{blockarray}{ccccccccc}
& &  \color{gray} \dots & \color{gray}{t(+e)} & \color{gray} \dots & \color{gray} s(+e) & \color{gray}\dots & \\
\begin{block}{cc (ccccccc)}
\color{gray} \vdots & &  \phantom{0} & \phantom{0} & \phantom{0} & \phantom{0} \\
\color{gray} t(+e) & & \phantom{0} & W_{-e} & \phantom{0} & W_{+e} & \phantom{0} & \phantom{0} \\
\color{gray} \vdots & &  \phantom{0} & \phantom{0} & \phantom{0} & \phantom{0} \\
\color{gray} s(+e) & & \phantom{0} & -W_{- e} & \phantom{0} & -W_{+e} & \phantom{0} \\
\color{gray} \vdots & &  \phantom{0} & \phantom{0} & \phantom{0} & \phantom{0}\\
\end{block}
\end{blockarray}\,\,.
\end{align}
As one can now realize, $\mathbb{D}_e=2 d_{S_e} \mathbb{W}$. We also note that $\boldsymbol{1}^\intercal \mathbb{D}_e=\boldsymbol{0}$ and thus $\boldsymbol{1}^\intercal \mathbb{D}_e \mathbb{W}^D \mathds{1}^{(n)} \boldsymbol{\pi}=0$. Consequently, Eq.~\eqref{eq:rhs-inversefrr-exp} becomes
\begin{align} \nonumber
&C^\mathfrak{m}_{ne}-W_{+e} C^\mathfrak{s}_{ns(+e)}+W_{-e} C^\mathfrak{s}_{nt(+e)}=-2 \boldsymbol{1}^\intercal \mathds{1}^{(n)} \mathbb{W}^D (d_{S_e} \mathbb{W}) \boldsymbol{\pi} \\ \ &=2\boldsymbol{1}^\intercal \mathds{1}^{(n)} d_{S_e} \boldsymbol{\pi}  =2 d_{S_e} \pi_n \,,
\end{align}
where in the second step we used Eq.~\eqref{eq:response-drazin}. This proves the second identity in \cref{eq:inverse-frr}. The first identity, $(\tau_e/j_e) d_{B_e} \pi_n=2 d_{S_e} \pi_n$, has been proven in Ref.~\cite{aslyamov2024nonequilibrium}. This concludes the proof of Eq.~\eqref{eq:inverse-frr}.

Equation~\eqref{eq:inverse-frr-cur} can be proven analogously. We have
\begin{align} \nonumber
&C^\mathfrak{j}_{ee'}-W_{+e'} C^\mathfrak{m}_{s(+e')e}+W_{-e'} C^\mathfrak{m}_{t(+e')e}\\ \nonumber
&=\delta_{ee'} \tau_e-\boldsymbol{1}^\intercal \mathbb{D}_{e'} \mathbb{W}^D \mathbb{J}_e \boldsymbol{\pi}-\boldsymbol{1}^\intercal \mathbb{J}_e \mathbb{W}^D \mathbb{D}_{e'} \boldsymbol{\pi} \\ \nonumber
&=\delta_{ee'} \tau_e-2 \boldsymbol{1}^\intercal \mathbb{J}_e \mathbb{W}^D (d_{S_{e'}} \mathbb{W}) \boldsymbol{\pi}=\delta_{ee'} \tau_e+2 \boldsymbol{1}^\intercal \mathbb{J}_e d_{S_{e'}} \boldsymbol{\pi} \\
&=\delta_{ee'} \tau_e+2W_{+e} d_{S_{e'}} \pi_{s(+e)}-2W_{-e} d_{S_{e'}} \pi_{t(+e)}=2 d_{S_{e'}} j_e \,,
\end{align}
where in the last step we used \cref{eq:current-expansion-asym}. This proves the second identity in \cref{eq:inverse-frr-cur}. The first identity, $(\tau_{e'}/j_{e'}) d_{B_{e'}} j_e=2d_{S_{e'}} j_e$, has been proven in Ref.~\cite{aslyamov2024nonequilibrium}.

\section{Derivation of Inverse FRRs~\eqref{eq:inverse-frr}--\eqref{eq:inverse-frr-cur} using Ref.~\cite{zheng2025spatial}} \label{app:proof-invfrr-zheng}
Here we present an alternative derivation of Inverse FRRs \eqref{eq:inverse-frr}--\eqref{eq:inverse-frr-cur} using the approach recently presented in Ref.~\cite{zheng2025spatial}, which can be regarded as a development of nonequilibrium FDTs from Refs.~\cite{agarwal1972fluctuation,baiesi2009fluctuations,seifert2010fluctuation}. Within this approach, $\hat{A}(t)$ is an arbitrary observable which is a function of the stochastic trajectory of the system during the time interval $[0,t]$. The time-integrated state and current observables defined via Eqs.~\eqref{eq:stateobsdef}--\eqref{eq:curobsdef} are  particular instances of such an observable. We then consider a dynamic response of that observable to a perturbation of transition rates $W_{\pm e}$ that starts at moment $t=0$. It can be determined as
\begin{align} \label{eq:zhengfrr}
d_{X_{\pm e}} \langle  \hat{A}(t) \rangle=\langle \Delta \hat{A}(t) [\Delta k_{\pm e}(t)-W_{\pm e} \theta_{s(\pm e)}(t)] \rangle \,,
\end{align}
where $X_{\pm e} \equiv \ln W_{\pm e}$, $\Delta \hat{A}(t) \equiv \hat{A}(t)-\langle \hat{A}(t) \rangle$, and the quantities $\theta_n(t)$ and $\Delta k_{\pm e}(t)$ were defined below \cref{eq:covmat-def}. This expression can be generalized to account for perturbations of any parameter $p$ parameterizing the transition rates $W_{\pm e}$, see Ref.~\cite{zheng2025spatial} for details.

To derive \cref{eq:inverse-frr}, we first use the chain rule for derivatives, $2 d_{S_e}\pi_n=d_{X_{+e}} \pi_n-d_{X_{-e}} \pi_n$. We then note that the static response corresponds to the infinite-time limit of dynamic responses, $d_{X_{+e}} \pi_n=\lim_{t\rightarrow \infty} t^{-1} \int_0^t dt' d_{X_{+e}} \langle \phi_n(t') \rangle$. Consequently, we obtain
\begin{widetext}
\begin{align}
&2 d_{S_e}\pi_n=d_{X_{+e}} \pi_n-d_{X_{-e}} \pi_n=\lim_{t\rightarrow \infty} t^{-1} \int_0^t dt' [d_{X_{+e}} \langle \phi_n(t') \rangle-d_{X_{-e}} \langle \phi_n(t') \rangle] \\ \nonumber
&=\lim_{t\rightarrow \infty} t^{-1} \langle \theta_n(t) [\Delta k_{+e}(t)-\Delta k_{-e}(t)-W_{+e} \theta_{s(+e)}(t)+W_{-e} \theta_{s(-e)}(t)] \rangle = C^\mathfrak{m}_{ne}-W_{+e} C^\mathfrak{s}_{ns(+e)}+W_{-e} C^\mathfrak{s}_{nt(+e)} \,.
\end{align}
\end{widetext}
Here, in the third step, we used \cref{eq:zhengfrr} and the definition of $\theta_n(t)$, while in the fourth step we used definitions from Eqs.~\eqref{eq:covmat-def}, \eqref{eq:covmat-def-state}, and the property $s(-e)=t(+e)$. This yields the second identity in \cref{eq:inverse-frr}. As noted in the Appendix~\ref{app:proof-frr}, the first identity, $(\tau_e/j_e) d_{B_e} \pi_n=2 d_{S_e} \pi_n$, has been proven in Ref.~\cite{aslyamov2024nonequilibrium}. 

The derivation of the second identity in \cref{eq:inverse-frr-cur} proceeds analogously,
\begin{widetext}
\begin{align}
&2 d_{S_{e'}} j_e=d_{X_{+e'}} j_e-d_{X_{-e'}} j_e=\lim_{t\rightarrow \infty} t^{-1} [d_{X_{+e'}} \langle k_{+e}(t)-k_{-e}(t) \rangle-d_{X_{-e'}} \langle k_{+e}(t)-k_{-e}(t) \rangle] \\ \nonumber
&=\lim_{t\rightarrow \infty} t^{-1} \langle [ \Delta k_{+e}(t)-\Delta k_{-e}(t)] [\Delta k_{+e'}(t)-\Delta k_{-e'}(t)-W_{+e'} \theta_{s(+e')}(t)+W_{-e'} \theta_{s(-e')}(t)] \rangle = C^\mathfrak{j}_{ee'}-W_{+e'} C^\mathfrak{m}_{s(+e')e}+W_{-e'} C^\mathfrak{m}_{t(+e')e} \,,
\end{align}
\end{widetext}
with the first identity in \cref{eq:inverse-frr-cur} proven in Ref.~\cite{aslyamov2024nonequilibrium}.

\section{Derivation of Eq.~\eqref{eq:frr-gen}} \label{app:der-frr-gen}
To derive Eq.~\eqref{eq:frr-gen}, we expand the state-flux covariance as
\begin{align} \label{eq:frr-gen-der}
\langle \!\langle \mathcal{O},\mathcal{F} \rangle \! \rangle=\sum_{n,\pm e} o_n y_{\pm e} \lim_{t \rightarrow \infty} t^{-1} \langle \theta_n(t) \Delta k_{\pm e}(t) \rangle \,.
\end{align}
Using Eq.~\eqref{eq:zhengfrr}, we obtain
\begin{align} \nonumber
&\lim_{t \rightarrow \infty} t^{-1} \langle \theta_n(t) \Delta k_{\pm e}(t) \rangle \\ \nonumber &=W_{\pm e} \lim_{t \rightarrow \infty} t^{-1} \langle \theta_n(t) \theta_{s(\pm e)}) \rangle+d_{X_{\pm e}} \lim_{t \rightarrow \infty} t^{-1} \int_0^t dt' \langle \phi_n(t') \rangle \\
&=W_{\pm e} C^\mathfrak{s}_{ns(\pm e)}+d_{X_{\pm e}} \pi_n \,.
\end{align}
We then use a version of FRR for state covariances derived in the companion letter~\cite{ptaszynski2024nonequilibrium},
\begin{align} 
C_{nm}^\mathfrak{s} &= \sum_{\pm e}\frac{1}{\varphi_{\pm e}}d_{X_{\pm e}} \pi_n d_{X_{\pm e}} \pi_m \,.
\end{align}
Inserting this into the expression above, we get
\begin{align} \nonumber
&\lim_{t \rightarrow \infty} t^{-1} \langle \theta_n(t) \Delta k_{\pm e}(t) \rangle \\ \nonumber &=W_{\pm e} \sum_{\pm e'} \frac{1}{\varphi_{\pm e'}} d_{X_{\pm e'}} \pi_n d_{X_{\pm e'}} \pi_{s(\pm e)}+d_{X_{\pm e}} \pi_n \\ &=\sum_{\pm e'} \frac{1}{\varphi_{\pm e'}} d_{X_{\pm e'}} \pi_n d_{X_{\pm e'}} \varphi_{\pm e} \,,
\end{align}
where in the last step we used $\varphi_{\pm e}=W_{\pm e} \pi_{s (\pm e)}$ and the chain rule of derivatives. Inserting this into Eq.~\eqref{eq:frr-gen-der}, and using Eqs.~\eqref{eq:stateobsmean} and~\eqref{eq:fluxobsdef}, we obtain Eq.~\eqref{eq:frr-gen}.

\section{Explicit expressions for $C_{\text{E},1b}^{\mathfrak{m}(e)}$} \label{app:exp-expr-ke1b}
The explicit expressions for the terms $C_{\text{E},1b}^{\mathfrak{m}(e)}$ defined in \cref{eq:ke1b} read
\begin{subequations}
\begin{align}
C_{\text{E},1b}^{\mathfrak{m}(1a)}&=-\frac{W_{-1a} W_{-2a}^2 W_{+1b} (2 W_{+1a}+W_{+1b}) (W_{-2a}+W_{+2a})}{[W_{-1a} W_{-2a}+ (W_{+1a}+W_{+1b})(W_{-2a}+W_{+2a})]^3} \,, \\
C_{\text{E},1b}^{\mathfrak{m}(1b)}&=\frac{W_{-1a} W_{-2a}^2 W_{+1b} [W_{-1a} W_{-2a}+W_{+1a} (W_{-2a}+W_{+2a})]}{[W_{-1a} W_{-2a}+ (W_{+1a}+W_{+1b})(W_{-2a}+W_{+2a})]^3} \,, \\
C_{\text{E},1b}^{\mathfrak{m}(2a)}&=\frac{2 W_{-1a} W_{-2a} W_{+1b} W_{+2a} (W_{+1a}+W_{+1b})^2}{[W_{-1a} W_{-2a}+ (W_{+1a}+W_{+1b})(W_{-2a}+W_{+2a})]^3} \,.
\end{align}
\end{subequations}

\bibliography{biblio}
\end{document}